\documentclass[sigconf]{acmart}

\copyrightyear{2022}
\acmYear{2022}
\setcopyright{acmcopyright}\acmConference[SEAMS '22]{17th International
Symposium on Software Engineering for Adaptive and Self-Managing Systems}{May
18--23, 2022}{Pittsburgh, PA, USA}
\acmBooktitle{17th International Symposium on Software Engineering for Adaptive
and Self-Managing Systems (SEAMS '22), May 18--23, 2022, Pittsburgh, PA, USA}
\acmPrice{15.00}
\acmDOI{10.1145/3524844.3528060}
\acmISBN{978-1-4503-9305-8/22/05}

\usepackage{pgfplots}
\usepackage{prettyref}
\usepackage{tikz}
\usepackage{xcolor,colortbl}
\usepackage{multirow}
\usepackage[ruled, linesnumbered, vlined, commentsnumbered]{algorithm2e}
\usepackage[section]{placeins}
\usepackage{booktabs}
\usepackage{threeparttable}
\usepackage{makecell}
\usepackage{graphicx}
\usepackage[listings,skins,breakable]{tcolorbox}
\usepackage{csquotes}
\usepackage{subcaption}
\usepackage{standalone}
\usepackage[skip=0.5\baselineskip]{caption}
\usepackage{wrapfig}
\usepackage{amsmath}
\usetikzlibrary{pgfplots.statistics}
\usepackage{tcolorbox}
\usepackage{pifont}
\usepackage{balance}
\usepackage{tablefootnote}

\newtcolorbox{rqbox}{colback=steel!10,boxrule=0.4pt,colframe=black,fonttitle=\bfseries,top=2pt,bottom=2pt}

\newtcolorbox{quotebox}{colback=green!10,boxrule=0.4pt,colframe=black,fonttitle=\bfseries,top=2pt,bottom=2pt}

\usepackage{pifont}
\newcommand{\cmark}{\textcolor{green!50!black}{\ding{51}}}%
\newcommand{\xmark}{\textcolor{red!80!black}{\ding{55}}}%

\SetKwInput{kwDeclare}{Declare}

\mathchardef\mhyphen="2D
\newcommand{\vect}[1]{\boldsymbol{#1}}
\DeclareMathAlphabet\mathbfcal{OMS}{cmsy}{b}{n}

\newbox\aMark
\setbox\aMark\hbox{\begin{pgfpicture}\textcolor{red}{\pgfuseplotmark{o}}\end{pgfpicture}}

\newbox\bMark
\setbox\bMark\hbox{\begin{pgfpicture}\textcolor{red}{\pgfuseplotmark{star}}\end{pgfpicture}}

\newrefformat{fig}{Fig.~\ref{#1}}
\newrefformat{tab}{Table~\ref{#1}}
\newrefformat{sec}{Section~\ref{#1}}
\newrefformat{app}{Appendix~\ref{#1}}
\newrefformat{alg}{Algorithm~\ref{#1}}
\newrefformat{property}{Property~\ref{#1}}
\newrefformat{theorem}{Theorem~\ref{#1}}
\newrefformat{lemma}{Lemma~\ref{#1}}
\newrefformat{corollary}{Corollary~\ref{#1}}
\newrefformat{proposition}{Proposition~\ref{#1}}
\newrefformat{def}{Definition~\ref{#1}}
\newrefformat{eq}{equation~(\ref{#1})}

\definecolor{steel}{rgb}{0, 0.2, 0.9} 

\pgfplotsset{compat=newest}

\pgfplotsset{compat=1.11,
    /pgfplots/ybar legend/.style={
    /pgfplots/legend image code/.code={%
       \draw[##1,/tikz/.cd,yshift=-0.25em]
        (0cm,0cm) rectangle (3pt,0.8em);},
   },
}

\DeclareMathOperator*{\argmin}{argmin}

\def\signed #1{{\leavevmode\unskip\nobreak\hfil\penalty50\hskip2em
  \hbox{}\nobreak\hfil(#1)%
  \parfillskip=0pt \finalhyphendemerits=0 \endgraf}}

\newsavebox\mybox

\begin{document}

\title{Planning Landscape Analysis for Self-Adaptive Systems}

\author{Tao Chen}
\affiliation{%
  \institution{Loughborough University}
  \state{Loughborough}
  \country{United Kingdom}}
  \email{t.t.chen@lboro.ac.uk}



\begin{abstract}

To assure performance on the fly, planning is arguably one of the most important steps for self-adaptive systems (SASs), especially when they are highly configurable with a daunting number of adaptation options. However, there has been little understanding of the planning landscape or ways by which it can be analyzed. This inevitably creates barriers to the design of better and tailored planners for SASs. In this paper, we showcase how the planning landscapes of SASs can be quantified and reasoned, particularly with respect to the different environments. By studying four diverse real-world SASs and 14 environments, we found that (1) the SAS planning landscapes often provide strong guidance to the planner, but their ruggedness and multi-modality can be the major obstacle; (2) the extents of guidance and number of global/local optima are sensitive to the changing environment, but not the ruggedness of the surface; (3) the local optima are often closer to the global optimum than other random points; and (4) there are considerable (and useful) overlaps on the global/local optima between landscapes under different environments. We then discuss the potential implications to the future work of planner designs for SASs. 

\end{abstract}

\maketitle

\section{Introduction}


Morden software systems are often engineered as highly configurable for handling different performance needs~\cite{ChenMMO21,DBLP:journals/corr/abs-2112-07303,DBLP:conf/sigsoft/JamshidiVKS18,DBLP:conf/kbse/LiXCT20,DBLP:conf/icse/LiX0WT20}, such as a good latency and throughput, even as the system runs~\cite{DBLP:conf/splc/LesoilATBJ21}. Therefore, self-adaptation --- the ability to find a better adaptation plan of the configurations that improves the performance on the fly --- is on high demand for highly configurable systems, which is the type of self-adaptive systems (SASs) we consider in this paper. 


For almost any type of SASs, planning is a key step in achieving self-adaptation and it mainly seeks to address one question: what is the best adaptation plan to take under a new environment (or when needed)? To this end, the community has relied on various different planning algorithms (or search algorithms), particularly the stochastic ones, to design a planner~\cite{DBLP:conf/sigsoft/ElkhodaryEM10,DBLP:journals/tse/CalinescuGKMT11,Chen2018FEMOSAA,DBLP:conf/sigsoft/ShahbazianKBM20,DBLP:journals/taas/KinneerGG21,DBLP:journals/tsc/ChenB17}. This equips a SAS with the ability to reason about the better or worse of the adaptation plans in the planning landscape, and hence ideally choose the one that has the best performance without the need to traverse the entire search space. From that regard, planning for SASs resembles a search and optimization process~\cite{DBLP:conf/saso/FredericksGK019}, which is complementary to many other techniques that are widely-used for SASs, such as control theory~\cite{DBLP:conf/icse/CamaraP0WGHT20}, machine learning~\cite{DBLP:conf/icse/Chen19b,DBLP:conf/icse/DoncktWQDM20,DBLP:conf/kbse/JamshidiSVKPA17,DBLP:journals/tse/ChenB17}, and formal verification~\cite{DBLP:conf/sigsoft/MandrioliM20}.

\begin{figure}[t!]
\centering
\begin{subfigure}[h]{0.52\columnwidth}
\includegraphics[width=\columnwidth]{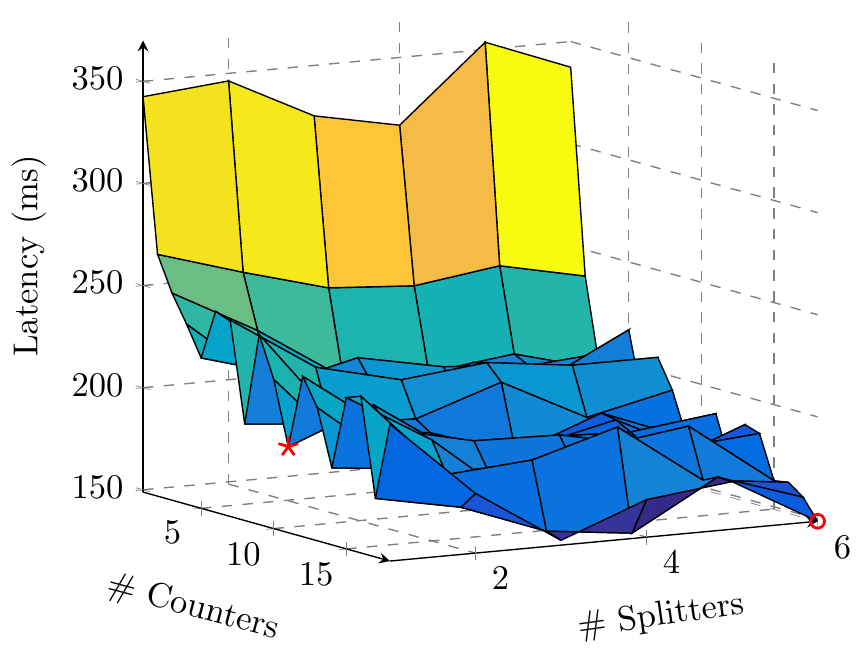}
\subcaption{\textsc{Storm} under \textsc{RollingCount}}
\end{subfigure}
~
\begin{subfigure}[h]{0.52\columnwidth}
\includegraphics[width=\columnwidth]{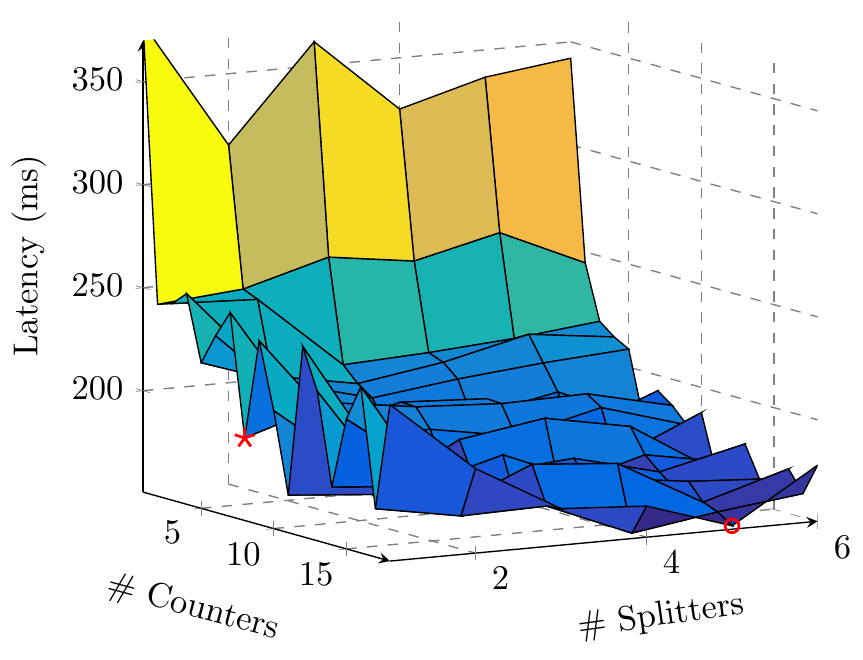}
\subcaption{\textsc{Storm} under \textsc{WordCount}}
\end{subfigure}
\caption{A projected landscape of the \textit{Latency} 
		with respect to adaptation options \texttt{Splitters} and \texttt{Counters} 
		for \textsc{Storm} under the \textsc{RollingCount} and \textsc{WordCount} environment.	
		\copy\aMark~and \copy\bMark~denote a global and local optimum, respectively.}
\label{fig:exp-land}
\end{figure}

Despite the importance, designing an effective planner is non-trivial, because the search space can be too large; the budget for a planner to conclude is limited; and there may be a complex relationship between adaptation plans and their performance, as well as the changing environment (e.g., there may be difficult local optima\footnote{Some sub-optimal points that have better or identical performance to all of their surrounding points.}). This becomes even more challenging when we have little understanding of the planning landscape of SASs or ways by which we can utilize to study it. Indeed, there have been some works on the planning algorithms for SAS~\cite{DBLP:conf/sigsoft/ElkhodaryEM10,DBLP:journals/tse/CalinescuGKMT11,Chen2018FEMOSAA,DBLP:conf/sigsoft/ShahbazianKBM20,DBLP:journals/taas/KinneerGG21,DBLP:journals/tsc/ChenB17}, it however remains unclear about what characteristics of the SAS planning landscape have enabled their success (or failure). 

Figure~\ref{fig:exp-land} shows an example of the projected planning landscapes for \textsc{Storm} under two environments. As can be seen, even with a clear visualization of such a simplified version (which is what has been commonly used in existing work~\cite{DBLP:conf/mascots/JamshidiC16,DBLP:conf/icse/DoncktWQDM20}), it is not always straightforward to quantify and conclude useful information/properties about the landscapes and their relative differences across environments. Of course, simply trying the planner on the SAS and examining the outcome is insufficient, even if we can do so. Rather, we need to be able to answer questions like whether the correlation between performance and plans in the landscape can provide strong guidance for a planner; how rugged the surface is; and what properties of the landscapes have been changed (or still unchanged) with the time-varying environment. We envisage that such a better understanding of the planning landscape is important, as it will enable us to derive planner designs that are explicitly tailored to the observed characteristics of SASs, promoting more effective and efficient planning which would otherwise be difficult.

To close the above gap, in this paper, we show how the metrics and notions from the domain of fitness landscape analysis~\cite{wright1932roles,DBLP:series/sci/PitzerA12} in the optimization community can be derived to study the planning landscape for SAS, particularly in relation to the different environments. This is achieved by an empirical analysis of four real-world SASs with 14 environments from the literature~\cite{ChenMMO21,DBLP:conf/mascots/MendesCRG20,DBLP:conf/icse/SiegmundKKABRS12,DBLP:conf/sigsoft/JamshidiVKS18,nair2018finding}, which are of different domains, languages, scales, and search spaces. The results reveal some interesting patterns:

\begin{itemize}
\item The SAS planning landscapes often provide useful information to guide the search process in a planner, but their ruggedness and multi-modality can pose a barrier.

\item The extents of guidance and number of global/local optima are sensitive to the changing environment, but not the ruggedness of the surface in the planning landscape.

\item Local optima are often closer to the global optimum than other random points.

\item Planning landscapes under different environments of a SAS often share a good amount of global/local optima. In particular, preserving the local optima of an environment into the newly changed one can be beneficial, as they may immediately become the global optimum therein. 

\end{itemize}

We then discuss the implications of our results for future planner design on SASs. To promote open science, we release the code and data in this work at: \textcolor{blue}{\url{https://doi.org/10.5281/zenodo.5866808}}.

In what follows, 
Section~\ref{sec:prob} introduces the background. 
Section~\ref{sec:landscape} elaborates on the notions and the metrics chosen for our landscape analysis. 
Section~\ref{sec:results} presents the methodology of study and analyzes the results. 
Key implications are discussed in Section~\ref{sec:discussion} and threats to validity are presented in Section~\ref{sec:threats}. 
Sections~\ref{sec:related} and~\ref{sec:con} analyze the related work and conclude the paper, respectively.

\section{Background}
\label{sec:prob}

When self-adapting highly-configurable systems, there are $n$ adaptation options such that the $i$th option is denoted as $x_i$, which can be a binary/integer variable.
The search space of all plans, $\mathbfcal{X}$, is the Cartesian product of the possible values or all the $x_i$. 

Without lose of generality, 
the ultimate goal of the SAS planning\footnote{We assume minimizing the performance objective.} is to achieve the following in a given environment:
\begin{align}
	\argmin~f(\vect{x}),~~\vect{x} \in \mathbfcal{X}
	\label{Eq:SOP}
\end{align}
where $\vect{x} = (x_1, x_2, ..., x_n)$ and $f$ measures the performance achieved by a plan, e.g., $\{3,1\}$ for the $\{$\texttt{num\_counters}, \texttt{num\_splitters}$\}$ on \textsc{Storm}. Since the environment can change as the SAS executes, the planning will run continuously. In the SAS literature, $f$ has been realized by different ways~\cite{DBLP:journals/csur/ChenBY18}, such as analytical models~\cite{DBLP:journals/ase/GerasimouCT18}, machine learning models~\cite{Chen2018FEMOSAA,DBLP:conf/icse/DoncktWQDM20,DBLP:conf/icse/ChenB13}, simulation~\cite{DBLP:conf/saso/FredericksGK019}, or even digital twins~\cite{DBLP:conf/acsos/AndersonWP21}, the details of which is outside the scope of this paper and hence we assume that there is a readily available resolution.


\begin{figure}[t!]
\centering
\includegraphics[width=0.8\columnwidth]{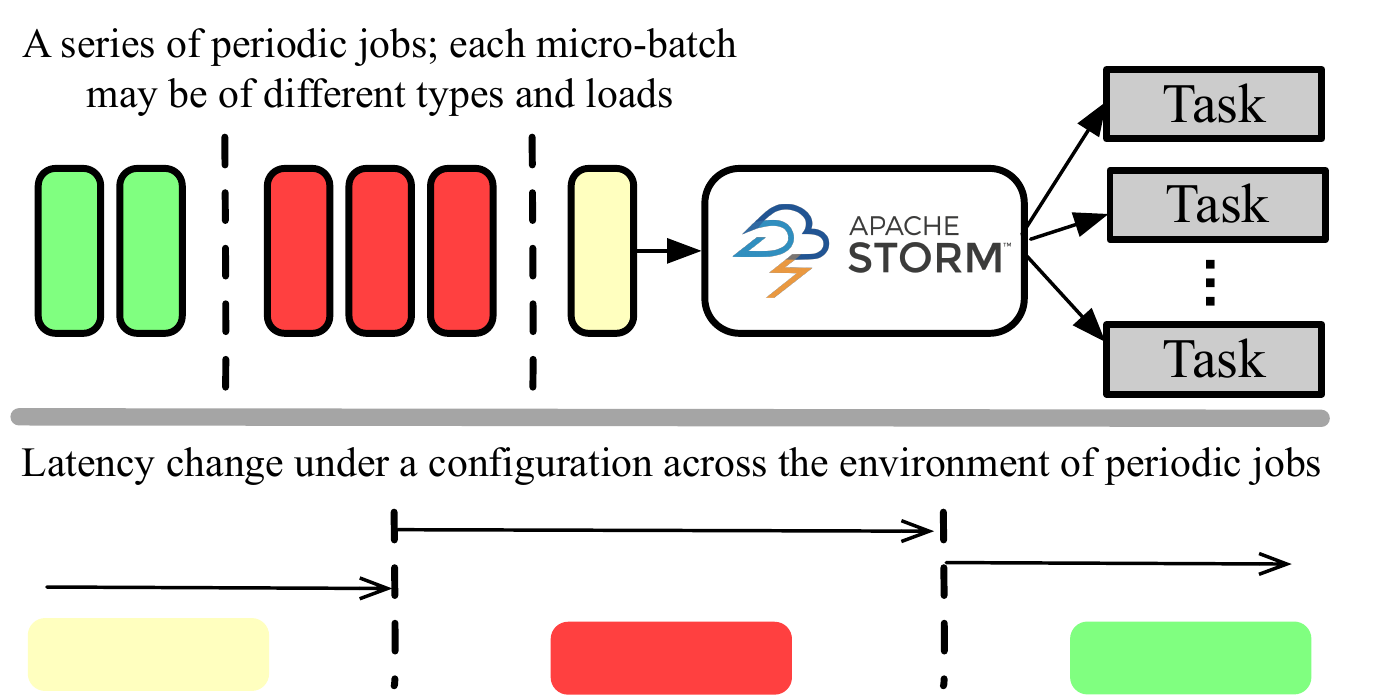}
\caption{Overview of \textsc{Storm} under changing environments.}
\label{fig:exp}
\end{figure}

Figure~\ref{fig:exp} illustrates \textsc{Storm}, which is a system that handles data streaming process under periodically incoming micro-batch of jobs (environments), e.g., \textsc{RollingCount} and \textsc{WordCount}. In this case, simply using a fixed configuration can be problematic: Jamshidi and Casale~\cite{DBLP:conf/mascots/JamshidiC16} have shown that using the default setting can lead to $480 \times$ slower than the best in some environments. This motivates the need for self-adaptation, where the aim is to optimize the latency by searching the right adaptation plan over changing environments.

\section{Landscape Analysis of SAS Planning}
\label{sec:landscape}

Fitness landscape is a concept initially coined by Wright~\cite{wright1932roles} and then extended to study the possible behaviours of algorithms in the optimization process, which fits precisely the needs of our planning analysis for SASs. In a nutshell, fitness landscape analysis concerns with understanding the relationships between the multi-dimensional encoding of the solutions (genotype) and their goodness (fitness) by means of various metrics and procedure~\cite{wright1932roles,DBLP:series/sci/PitzerA12}. 




Formally, the landscape for SAS planning under an environment can be represented as a tuple $\mathbfcal{F} = (\mathbfcal{X}, f, \mathbfcal{N}_k) $, such that:
\begin{align}
	 \mathbfcal{N}_k(x) = \{y \in \mathbfcal{X} : D(x,y) \leq k\}
\end{align}
whereby $\mathbfcal{X}$ is a set of points (adaptation plans); $f$ is the same performance table as that in Equation~\ref{Eq:SOP}. $\mathbfcal{N}_k$ is the neighborhood defined over set $\mathbfcal{X}$ according to a distance metric $D$ of size $k$ (which may be the bound for some operators to transform one plan into another). Clearly, when $k$ covers all the neighboring plans in the search space, we obtain a complete planning landscape for the SAS.

\subsection{Distance, Neighborhood and Local Optima}
\label{sec:h-dis}

Quantifying the distance between adaptation plans (and hence the neighborhood) is the fundamental step in our planning landscape analysis. In this work, we use Hamming distance $D_H$ to measure two adaptation plans because of three reasons:


\begin{itemize}


\item The sparse comparison in Hamming distance fits well with the landscape nature of configurable system~\cite{nair2018finding,DBLP:conf/mascots/JamshidiC16,ChenMMO21}.


\item Hamming distance does not quantify the magnitude of difference on an adaptation option, which fits with the need of most categorical options in configurable systems.

\item It is widely used in many real-world problems~\cite{DBLP:journals/tsmc/TavaresPC08,DBLP:journals/tec/MerzF00,DBLP:conf/gecco/OchoaQB09}.
\end{itemize}

\begin{figure}[t!]
\centering
\includegraphics[width=0.6\columnwidth]{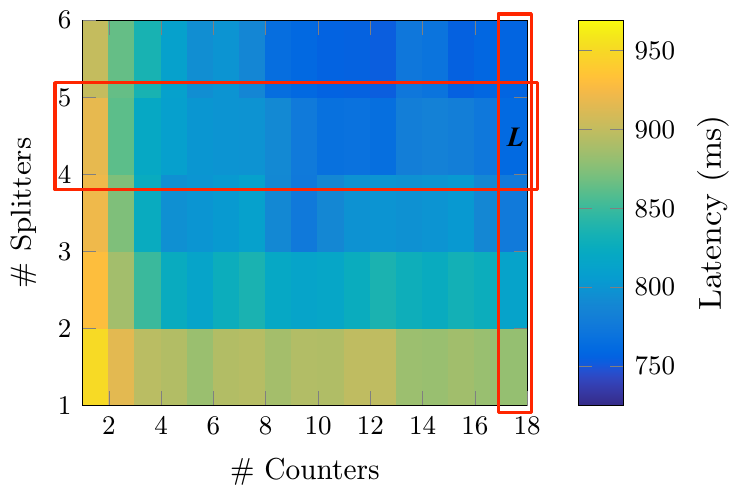}
\caption{Heatmap of a projected planning landscape for the \textit{Latency} of \textsc{Storm}. The block ``L'' is a local optimum under the neighborhood radius of one option (in red frames).}
\label{fig:lo}
\end{figure}

In this work we set $D_H =1$ hence the neighbors are the plans that differ exactly on one option. A point is a local optimum if it is no worse than all of its neighbors, as shown in Figure~\ref{fig:lo}.

\subsection{Fitness Distance Correlation}

Generally speaking, Fitness Distance Correlation (FDC) examines how close is the relation between fitness value and distance to the nearest optimum in the search space~\cite{DBLP:conf/icga/JonesF95}, which quantifies the overall guidance that the planning landscape can offer for a planner~\cite{DBLP:journals/tsmc/TavaresPC08}. Formally, FDC (denoted as $\varrho$) is computed as:
\begin{align}
	 \varrho(f,d) = {1 \over {\sigma_f \sigma_d p}} \sum^p_{i=1} (f_i - \overline{f}) (d_i - \overline{d})
	 \label{Eq:fdc}
\end{align}
where $p$ is the number of points considered in FDC; in this work, we set $p$ as the total number of adaptation plans in the space and hence the FDC reflects the complete planning landscape. $f_i = f(x_i)$ is the performance value for the $i$th adaptation plan and $d_i = d_{opt}(x_i)$ is the shortest Hamming distance of such a plan to a global optimum. $\overline{f}$ ($\overline{d}$)
and $\sigma_f$ ($\sigma_d$) are the mean and standard deviation, respectively.

Intuitively, FDC is in fact the Pearson correlation between $f$ and $d$, hence it ranges on $[-1,1]$ where $1$ and $-1$ imply the strongest monotonically positive and negative correlation, respectively; $0$ indicates no correlation can be detected. Since in our case we prefer a smaller performance value, when $0 < \varrho \leq 1$, the adaptation plan turns better (smaller performance value) as the shortest distance to a global optimum reduces. This means that, when FDC becomes closer to $1$, the guidance provided to a planner is stronger and it is more likely to exist a path towards a global optimum via adaptation plans with decreasing performance values, hence the planning can be reasonably solved. In contrast, $-1 \leq \varrho < 0$ indicates the opposite. 


\subsection{Landscape Structure}
\label{sec:correlation}


In this work, we also explicitly assess the structure of the planning landscape, i.e., multi-modality and ruggedness. 


\subsubsection{Multi-modality}

In general, the multi-modality, as opposed to the uni-modality with one global optimum and no local optimum, refers to a special property of the landscape where there is more than one global/local optimum~\cite{DBLP:conf/foga/HornG94}. To quantify such, we can count the percentage of global/local optima in the complete planning landscape as a global metric of its structure~\cite{DBLP:conf/gecco/MersmannBTPWR11,DBLP:conf/foga/HornG94}. A landscape with a high degree of multi-modality is an indication that it contains many ``troughs'' (for minimizing objectives), which may both be a challenge and an opportunity. On one hand, multi-modality implies a complex structure (at least globally) and hence can raise the additional difficulty for the planner. On the other hand, the presence of different ``troughs'', together with the ability to locate them, can be beneficial for a planner to eventually reach a global optimum.

\subsubsection{Ruggedness}

Measuring multi-modality by the number of global/local optima still cannot account for the local paths between local optima, and the other related points. As a result, we additionally measure the Correlation Length ($\ell$) of the landscape~\cite{stadler1996landscapes} --- a local metric that indicates the local ruggedness property. 

To be more specific, the Correlation Length is the results of randomly sampled adaptation plans in the landscape, and hence it models the local surface of traversal that a planner would likely to explore. Formally, $\ell$ is calculated as below:
\begin{align}
		\ell(p,s) = - (\ln|{1 \over {\sigma_f^2 (p-s)}} \sum^{p-s}_{i=1} (f_i - \overline{f}) (f_{i+s} - \overline{f})|)^{-1} 
		 \label{Eq:rugg}
\end{align}
$\ell(p,s)$ is essentially a nomralized autocorrelation function of neighboring points' performance values explored and the notations are the same as that for Equation~\ref{Eq:fdc}. In this work, we conduct sampling with random walk~\cite{DBLP:journals/tsmc/TavaresPC08}, thus $s$ denotes the step size and $p$ is the walk length. We use $s=1$ in this work, which means that we target the most restricted form where the autocorrelation is calculated on adaptation plans sampled from adjacent steps (Note that the correlation cannot be $0$), as this is what has been widely followed~\cite{DBLP:journals/tsmc/TavaresPC08,DBLP:journals/tec/MerzF00,DBLP:conf/gecco/OchoaQB09}. The higher the value of $\ell$, the smoother the landscape, as the performance of adjacently sampled adaptation plans are more correlated. Otherwise, it indicates a more rugged surface~\cite{stadler1996landscapes}, which means the easier to trap a planner.

\section{Methodology and Results}
\label{sec:results}

In this work, we seek to answer the following research questions:

\begin{itemize}

\item \textbf{RQ1:} Do the planning landscapes offer useful guidance to a SAS planner under different environments?

\item \textbf{RQ2:} What are the general structural properties of planning landscapes for SASs under different environments?

\item \textbf{RQ3:} Do the local optima closer to global optimum than other non-optimal points in the SAS planning landscape?

\item \textbf{RQ4:} Is it possible to share some information on the planning landscapes for SAS across different environments?

\end{itemize}

To that end, as shown in Table~\ref{tb:sys}, we consider 3-4 environments that can change arbitrarily at runtime for four real-world SAS and use the same setting as previous work. We exploit the readily available dataset of those systems~\cite{ChenMMO21,nair2018finding,DBLP:conf/mascots/JamshidiC16,DBLP:conf/sigsoft/JamshidiVKS18,DBLP:conf/mascots/MendesCRG20} which contain the samples for the entire planning landscape of each environment.

\begin{table}[t!]
\caption{Real-world self-adaptive systems and their environments studied. We use the deep neural network in \textsc{Keras}. Reference shows the work that also uses the same systems.}
\label{tb:sys}
\setlength{\tabcolsep}{1mm}
\centering
\footnotesize

  \begin{center}
   \begin{adjustbox}{max width = 1\textwidth}
 \begin{threeparttable}
\begin{tabular}{lllcc}\toprule

\textbf{Subject SAS}&\textbf{Performance}&\textbf{Environments}&\textbf{$\#$ Options}&\textbf{Search Space}\\

\midrule

\multirow{4}{*}{\textsc{Storm~\cite{ChenMMO21,DBLP:conf/mascots/JamshidiC16}}}&\multirow{4}{*}{Latency}&$E_1$: \textsc{Speed-Of-Light}&\multirow{4}{*}{12}&\multirow{4}{*}{2,048}\\

&&$E_2$: \textsc{RollingSort}&&\\ 
&&$E_3$: \textsc{WordCount}&&\\ 
&&$E_4$: \textsc{RollingCount}&&\\ \hline

\multirow{4}{*}{\textsc{Keras}~\cite{DBLP:conf/sigsoft/JamshidiVKS18,DBLP:conf/mascots/MendesCRG20}}&\multirow{4}{*}{Inferred Time}&$E_1$: \textsc{ShapesAll}&\multirow{4}{*}{12}&\multirow{4}{*}{4,096}\\

&&$E_2$: \textsc{DSR}&&\\ 
&&$E_3$: \textsc{Adiac}&&\\ 
&&$E_4$: \textsc{Coffee}&&\\ \hline

\multirow{3}{*}{\textsc{x264}~\cite{ChenMMO21,nair2018finding}}&\multirow{3}{*}{Runtime}&$E_1$: \textsc{8MB}&\multirow{3}{*}{16}&\multirow{3}{*}{4,000}\\

&&$E_2$: \textsc{32MB}&&\\ 
&&$E_3$: \textsc{128MB}&&\\ \hline

\multirow{3}{*}{\textsc{Spear}~\cite{DBLP:conf/icse/SiegmundKKABRS12}}&\multirow{3}{*}{Runtime}&$E_1$: \textsc{4435 CXTY}&\multirow{3}{*}{14}&\multirow{3}{*}{16,384}\\

&&$E_2$: \textsc{8827 CXTY}&&\\
&&$E_3$: \textsc{10286 CXTY}&&\\


\bottomrule
\end{tabular}
 \end{threeparttable}
  \end{adjustbox}
     \end{center}
\end{table}

For interpreting the FDC and Correlation Lengths (\textbf{RQ1} and \textbf{RQ2}), we adopt Fisher's transformation~\cite{fisher1915frequency} to find the z-score, which is then interpreted using Zou's confidence interval~\cite{zou2007toward} under a significance level of $0.05$. We leverage the non-paired Wilcoxon rank-sum test at $\alpha=0.05$ for comparing the distance in \textbf{RQ3}.

\subsection{RQ1: Fitness Guidance in Planning}

\subsubsection{Method}

To answer \textbf{RQ1}, we leverage the FDC coefficient to measure the extents of guidance that a planning landscape offers to the planner, considering all SASs and their environments studied. To further interpret the FDC coefficients in detail, we adopt the classification concluded by Jones and Forrest~\cite{DBLP:conf/icga/JonesF95} (the values are reversed as we focus on minimizing the performance objectives):

\begin{itemize}


\item \textbf{Misleading ($\varrho \leq -0.15$).} The landscape can drive the search to move away from the global optimum.

\item \textbf{Difficult ($-0.15 < \varrho < 0.15$).} The correlation is insignificant to guide the planner on any particular direction.

\item \textbf{Straightforward ($\varrho \geq 0.15$).} The landscape provides useful guidance for a planner to reach a global optimum.

\end{itemize}


\begin{figure}[t!]
\centering

\begin{subfigure}[h]{0.8\columnwidth}
\includegraphics[width=\columnwidth]{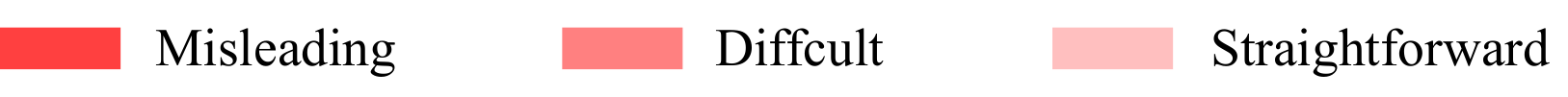}
\end{subfigure}

\begin{subfigure}[h]{0.25\columnwidth}
\includegraphics[width=\columnwidth]{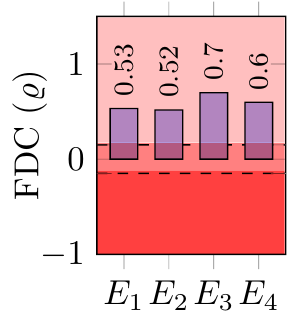}
\subcaption{\textsc{Storm}.}
\end{subfigure}
~
\begin{subfigure}[h]{0.25\columnwidth}
\includegraphics[width=\columnwidth]{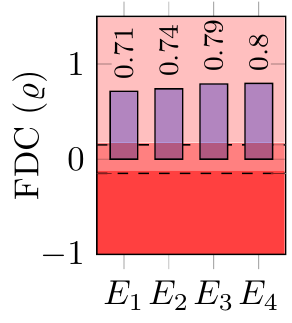}
\subcaption{\textsc{Keras}.}
\end{subfigure}
~
\begin{subfigure}[h]{0.25\columnwidth}
\includegraphics[width=\columnwidth]{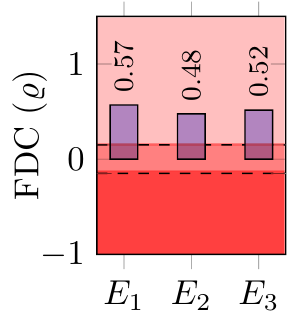}
\subcaption{\textsc{x264}.}
\end{subfigure}
~
\begin{subfigure}[h]{0.25\columnwidth}
\includegraphics[width=\columnwidth]{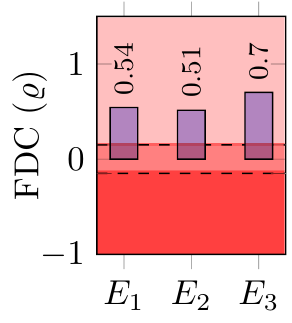}
\subcaption{\textsc{Spear}.}
\end{subfigure}

\caption{The FDC coefficients on all SASs and environments.}
\label{fig:rq1}
\end{figure}

\subsubsection{Result}

The FDC coefficients have been plotted in Figure~\ref{fig:rq1} with statistical test results between each pair of environments for a SAS in Table~\ref{tb:rq1}. As can be seen, we obtain a few interesting findings:

\begin{enumerate}

\item Surprisingly, the planning landscapes on all SASs and environments are classified as \textit{``Straightforward''}, implying that they offer a good degree of useful guidance for the planner.

\item The FDC coefficients tend to differ across the SASs (with the landscapes of \textsc{Keras} showing the strongest guidance), which is as expected since they all come from different domains and are implemented in different languages.

\item Although the FDC coefficients under different environments of a SAS may be seen as similar, most of their differences are statistically significant, i.e., in 15 out of 18 cases (Table~\ref{tb:rq1}).

\end{enumerate}

Therefore, for \textbf{RQ1}, we say:

\begin{quotebox}
 \noindent
 \textit{\textbf{To RQ1:} Yes, the planning landscapes offer useful and strong guidance to the SAS planner in general, but the environmental change can influence the guidance provided.}
\end{quotebox}

\subsection{RQ2: Planning Landscape Structure}

\subsubsection{Method}

The structure in \textbf{RQ2} is measured in two ways: (1) the \% of global/local optima in the landscape (for multi-modality) and (2) the Correlation Length (for ruggedness), as discussed in Section~\ref{sec:correlation}, for which we set the random walk length of 50 ($p=50$ from Equation~\ref{Eq:rugg}) with 50 repeats and report the mean values.



While we cannot find a general standard to classify the degree of ruggedness similar to that for FDC, to aid our interpretation, we turned into the literature from the general optimization community. We hence use the calculated Correlation Length for some common problems with well-acknowledged challenges on local optima as the baselines in our discussion\footnote{Those problems can have different instances of the landscape; in this work, we use the smallest mean Correlation Length for each (most rugged surface) as reported.}, such as Multidimensional Knapsack Problem (MKP)~\cite{DBLP:journals/tsmc/TavaresPC08}, Quadratic Assignment Problem (QAP)~\cite{DBLP:journals/tec/MerzF00}, and Timetabling Problem (TP)~\cite{DBLP:conf/gecco/OchoaQB09}. This is possible as Correlation Length is a scale- and unit-agnostic metric.

\begin{table}[t!]
\caption{Statistical test ($p$ value) on the FDC coefficient ($\varrho$) and Correlation Length ($\ell$) between all pairs of environments. Statistically significant ones are highlighted in bold.}
    \label{tb:rq1}
   \footnotesize
    \setlength{\tabcolsep}{1mm}
  \begin{center}
    \begin{adjustbox}{max width = 1\columnwidth}

    \begin{tabular}{c|cccc||cccc} \toprule
    \multicolumn{1}{c}{}&\multicolumn{4}{c}{\textbf{FDC coefficient ($\varrho$)}}&\multicolumn{4}{c}{\textbf{Correlation Length ($\ell$)}}\\ \hline
    \textbf{}&\textsc{Storm}&\textsc{Keras}&\textsc{x264}&\textsc{Spear}&\textsc{Storm}&\textsc{Keras}&\textsc{x264}&\textsc{Spear}\\ \hline
    
    $E_1 : E_2$&0.4742&\textbf{0.0215}&\textbf{$<$0.0001}&\textbf{0.0004}&0.7340&0.8712&0.8802&0.9142 \\
    $E_1 : E_3$&\textbf{$<$0.0001}&\textbf{$<$0.0001}&\textbf{0.0151}&\textbf{$<$0.0001}&\textbf{0.029}&0.8095&0.6799&0.9757 \\
    $E_1 : E_4$&\textbf{0.0021}&\textbf{$<$0.0001}&N/A&N/A&0.0516&0.8012&N/A&N/A \\
    $E_2 : E_3$&\textbf{$<$0.0001}&\textbf{$<$0.0001}&0.0971&\textbf{$<$0.0001}&0.0659&0.9370&0.7934&0.8900 \\
    $E_2 : E_4$&\textbf{0.0002}&\textbf{0.0001}&N/A&N/A&0.1081&0.9285&N/A&N/A \\
    $E_3 : E_4$&\textbf{$<$0.0001}&0.4454&N/A&N/A&0.8163&0.9914&N/A&N/A \\

    \bottomrule
    \end{tabular}
  \end{adjustbox}
   \end{center}
\end{table}

\begin{figure}[t!]
\centering

\begin{subfigure}[h]{0.48\columnwidth}
\includegraphics[width=\columnwidth]{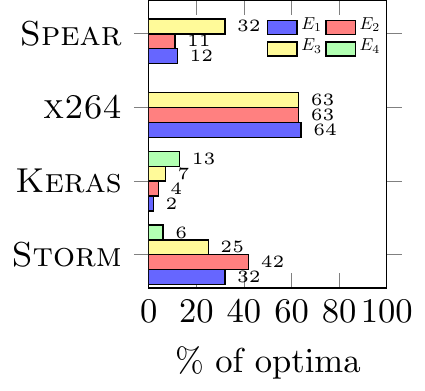}
\subcaption{\% of the global/local optima.}
\end{subfigure}
~
\begin{subfigure}[h]{0.52\columnwidth}
\includegraphics[width=\columnwidth]{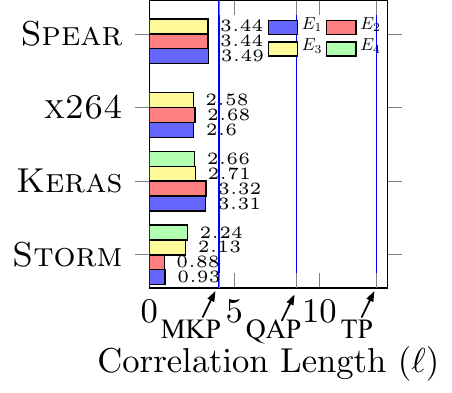}
\vspace{-0.45cm}
\subcaption{Mean $\ell$ over 50 repeats.}
\end{subfigure}

\caption{The landscape ruggedness on all SASs and environments. The blue lines in (b) are the smallest $\ell$ (most rugged landscape) of the other widely-studied problems.}
\label{fig:rq2}
\end{figure}

\subsubsection{Result}

From the results in Figure~\ref{fig:rq2} and the statistical tests of $\ell$ between environments in Table~\ref{tb:rq1}, we observe that:

\begin{enumerate}

\item The percentage of global/local optima in SAS planning landscape indicates a reasonable sign of multi-modality in general. For some SASs, such as \textsc{x264}, it can go over 50\%.

\item Different SASs and their environments often significantly affect the percentage of global/local optima in the planning landscape, but the overall multi-modal property is unaffected. The planning landscapes of \textsc{x264} exhibit a much higher degree of multi-modality than the other SASs and appears to be insensitive to environmental change.

\item The Correlation Lengths of all SASs/environments are lower than that of MKP (which has the smallest value), suggesting that the ruggedness of SAS planning landscape is non-trivial.

\item The Correlation Length differs considerably across the SASs (with landscapes for \textsc{Storm} being the most rugged ones), but between the environments for the same SAS, the differences are often insignificant. This has also been evidenced in Table~\ref{tb:rq1} where only one case has $p<0.05$.

\end{enumerate}

In summary, we answer \textbf{RQ2} as:

\begin{quotebox}
 \noindent
 \textit{\textbf{To RQ2:} The planning landscapes for SASs show a good sign of multi-modality and they are more rugged than some other widely-studied problems. Yet, the ruggedness is insensitive to the changing environment but the multi-modality does.}
\end{quotebox}

\subsection{RQ3: Distance to Global Optimum}

\begin{figure}[t!]
\centering

\begin{subfigure}[h]{0.8\columnwidth}
\includegraphics[width=\columnwidth]{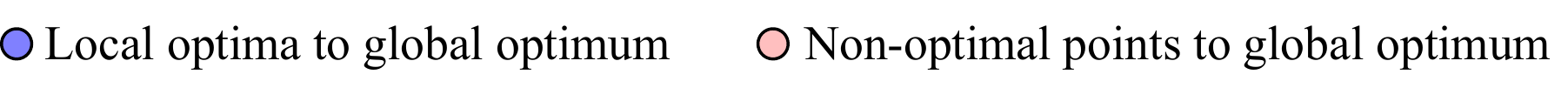}
\end{subfigure}

\begin{subfigure}[h]{0.25\columnwidth}
\includegraphics[width=\columnwidth]{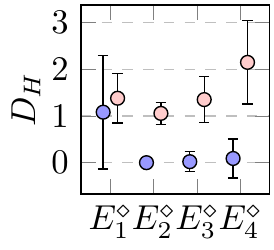}
\subcaption{\textsc{Storm}.}
\end{subfigure}
~
\begin{subfigure}[h]{0.25\columnwidth}
\includegraphics[width=\columnwidth]{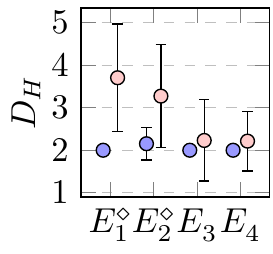}
\subcaption{\textsc{Keras}.}
\end{subfigure}
~
\begin{subfigure}[h]{0.25\columnwidth}
\vspace{-0.08cm}
\includegraphics[width=\columnwidth]{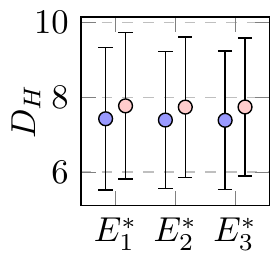}
\vspace{-0.4cm}
\subcaption{\textsc{x264}.}
\end{subfigure}
~
\begin{subfigure}[h]{0.25\columnwidth}
\includegraphics[width=\columnwidth]{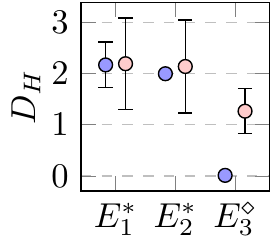}
\subcaption{\textsc{Spear}.}
\end{subfigure}

\caption{The shortest distance (mean and standard deviation) from the global optimum to local optima and the rest non-optimal ones, over all SASs and environments studied. $^{*}$ and $^{\diamond}$ denote $10^{-4} \leq p < 0.05$ and $p<10^{-4} $, respectively.}
\label{fig:rq3}
\end{figure}

\subsubsection{Method}

To study \textbf{RQ3}, we report on the overall Hamming distance ($d_{local}$) between local optima and their closest global optimum, together with that ($d_{others}$) between the rest of non-optimal points and the corresponding closet global optimum. 


\subsubsection{Result}

As we can see from Figure~\ref{fig:rq3}, the overall distances of $d_{local}$ is clearly shorter than $d_{others}$, regardless of the SASs and their environments; the extents of difference differ depending on the environment for a SAS though. The comparisons also come with statistical significance in general ($p<0.05$). This is strong evidence that the local optima in the planning landscapes, although may trap the planner, can be useful for serving as the ``stepping stones'' which may eventually lead to a global optimum.

Therefore, for \textbf{RQ3}, we say:

\begin{quotebox}
 \noindent
 \textit{\textbf{To RQ3:} The local optima are indeed generally closer to the global optimum than the other non-optimal points in SAS planning landscape, meaning that preserving then jumping out from them is more likely to reach a global optimum.}
\end{quotebox}

\begin{table}[t!]
\caption{The patterns on the global and local optima of adaptation plans shared between different environments of the SASs. $E_x$$\rightarrow$$E_y$ means the environment changes from $E_x$ to $E_y$. For $A_1$ and $A_2$, \cmark and \xmark~denote ``Yes'' and ``No'', respectively.}
    \label{tb:rq4}
    \footnotesize
    \setlength{\tabcolsep}{1.2mm}
  \begin{center}
    \begin{adjustbox}{max width = 1\textwidth}

    \begin{tabular}{cccc|ccc|ccc|ccc} \toprule
    \textbf{}& \multicolumn{3}{c|}{\textsc{Storm}}& \multicolumn{3}{c|}{\textsc{Keras}}& \multicolumn{3}{c|}{\textsc{x264}}& \multicolumn{3}{c}{\textsc{Spear}}\\ \cmidrule{2-13}
    
   \textbf{}&$A_1$&$A_2$&$A_3$&$A_1$&$A_2$&$A_3$&$A_1$&$A_2$&$A_3$&$A_1$&$A_2$&$A_3$\\ \midrule
    
 $E_1$$\rightarrow$$E_2$&\cmark&\cmark&56\%&\cmark&\cmark&31\%&\xmark&\cmark&85\%&\cmark&\cmark&18\% \\   
 
 $E_2$$\rightarrow$$E_1$&\cmark&\xmark&44\%&\cmark&\cmark&13\%&\xmark&\cmark&86\%&\cmark&\xmark&19\% \\   
  
 $E_1$$\rightarrow$$E_3$&\xmark&\cmark&38\%&\cmark&\cmark&95\%&\xmark&\cmark&85\%&\cmark&\cmark&47\% \\   
 
 $E_3$$\rightarrow$$E_1$&\xmark&\cmark&50\%&\cmark&\xmark&20\%&\xmark&\cmark&87\%&\cmark&\xmark&17\% \\ 
 
 $E_1$$\rightarrow$$E_4$&\xmark&\cmark&10\%&\cmark&\cmark&100\%&N/A&N/A&N/A&N/A&N/A&N/A \\ 
 
 $E_4$$\rightarrow$$E_1$&\xmark&\xmark&49\%&\cmark&\xmark&11\%&N/A&N/A&N/A&N/A&N/A&N/A  \\ 
 
 $E_2$$\rightarrow$$E_3$&\cmark&\xmark&38\%&\cmark&\cmark&63\%&\cmark&\xmark&86\%&\cmark&\cmark&65\% \\ 
 
 $E_3$$\rightarrow$$E_2$&\cmark&\cmark&64\%&\cmark&\xmark&33\%&\cmark&\xmark&88\%&\cmark&\xmark&23\% \\ 
 
 $E_2$$\rightarrow$$E_4$&\cmark&\xmark&11\%&\cmark&\cmark&100\%&N/A&N/A&N/A&N/A&N/A&N/A  \\ 
 
 $E_4$$\rightarrow$$E_2$&\cmark&\cmark&72\%&\cmark&\xmark&28\%&N/A&N/A&N/A&N/A&N/A&N/A  \\ 
 
 $E_3$$\rightarrow$$E_4$&\cmark&\xmark&25\%&\cmark&\cmark&100\%&N/A&N/A&N/A&N/A&N/A&N/A  \\ 
 
 $E_4$$\rightarrow$$E_3$&\cmark&\cmark&98\%&\cmark&\xmark&53\%&N/A&N/A&N/A&N/A&N/A&N/A  \\ 
  

    \bottomrule
    \end{tabular}
  \end{adjustbox}
   \end{center}
\end{table}


\subsection{RQ4: Information Between Environments}

\subsubsection{Method}

In \textbf{RQ4}, we study three aspects to understand the information of the planning landscapes that can be shared between environments, e.g., when changing from environment $E_x$ to $E_y$: 

\begin{itemize}

\item $A_1$: Whether there is at least one global optimum in $E_x$ that is also a global optimum under $E_y$.


\item $A_2$: Whether there is at least one local optimum in $E_x$ that would become a global optimum under $E_y$.

\item $A_3$: The percentage of global/local optima in $E_x$ that are also global/local optima under $E_y$.

\end{itemize}

\subsubsection{Result}

We demonstrate the results in Table~\ref{tb:rq4}, from which we can disclose some interesting observations:

\begin{enumerate}

\item From $A_1$ and $A_2$, there is only one case ($E_4 \rightarrow E_1$ for \textsc{Storm}) where the answers for both $A_1$ and $A_2$ are ``No''. This, together with the findings for \textbf{RQ3}, suggests that either the global or local optima (sometimes both) in one environment often helps the planner to find a global optimum in another.

\item From $A_1$ and $A_3$, we see that generally, the global optimum in one environment can directly serve as the global optimum after the environment change, as in 28 out of 36 cases there is at least one global optimum that satisfies $A_1$ (with \textsc{Keras} and \textsc{Spear} having overlapped global optimum over all environments.) and over half of the cases with more than 50\% global/local optima overlap. \textsc{x264} also exhibit particularly high overlapping between environments.

\item From $A_2$ and $A_3$ under a considerable overlap of global/local optima, there are 21 out of 36 cases in which at least one local optimum of an environment would become a global optimum after changing to another. This implies that preserving the local optima can be useful for a new environment.

\end{enumerate}

At this point, we conclude \textbf{RQ4} as:

\begin{quotebox}
 \noindent
 \textit{\textbf{To RQ4:} Yes, preserving both the global and local optima of the planning landscape in one environment can be useful for SAS planning under a changing environment.}
\end{quotebox}

\section{Implications}
\label{sec:discussion}

%
%
%
%
%
%
%
%
%
%
%
%

Our findings can excite a few research directions for SAS planning.

\textbf{RQ1} suggests that, for future research on \textbf{selecting proper planning algorithm}, SAS planning landscape is suitable for those algorithms guided by the fitness (performance), and jump from one adaptation plan to another by taking the distance to the currently reached adaptation plans into account, e.g., change one (or some more) option each time such as the GA with neighborhood-based mutation~\cite{wright1932roles,DBLP:journals/tsmc/TavaresPC08}. In contrast, a planner that alters the adaptation plan in a random number of options each time will likely lose the valuable guidance of the landscape. When exploited fully, such guidance is not only useful when the search space is intractable, but also helps to find global optimum quicker in a tractable space~\cite{wright1932roles}, which is attractive for SAS planning. However, it is important to inspect the impact caused by environmental change.


For \textbf{planner component design} from \textbf{RQ2}, we show that a mechanism which helps the planner to escape from local optima is indeed necessary. These mechanisms are readily available, such as larger radius of changes~\cite{DBLP:conf/sigmetrics/YeK03}, random restarting~\cite{DBLP:conf/hpdc/LiZMTZBF14}, accepting inferior plans~\cite{DBLP:conf/icpads/DingLQ15}, and multi-objectivization~\cite{ChenMMO21}. Albeit the number of local optima from the landscapes may differ, a mechanism that works for one environment will likely work for the others too. 


Finally, from \textbf{RQ3} and \textbf{RQ4}, we confirm again the importance of \textbf{seeded planning and plan reuse} for SASs under changing environments, which has recently attracted attention in the community~\cite{DBLP:journals/taas/KinneerGG21,DBLP:journals/infsof/ChenLY19,DBLP:conf/gecco/0001LY18}. Additionally, we provide initial evidence on what should be seeded and shared. The most surprising finding is that the local optima of an environment are also helpful for the planning under the new environment (in addition to the global optimum), as they may still be the local optima or one of them may immediately become the new global optimum therein. This, together with the finding that local optima are very much closer to the global optimum than other random points and hence more helpful (if the planner can escape from them), raises an interesting topic of multi-modal planning for SAS: in addition to finding the global optimum, we are also interested in preserving as many local optima as possible during a planning run~\cite{ChenLiDOS22}.

\section{Threats to Validity}
\label{sec:threats}


Threats to \textbf{internal validity} can be related to the setting of one option for defining local optimum and the step of 50 for the random walking that computes Correlation Length, which were decided pragmatically based on the needs. Larger values may change the absolute figures but are unlikely to invalidate the conclusion.


The metrics and evaluation used may possess threats to \textbf{construct validity}. In this work, the most common metrics from fitness landscape analysis are used~\cite{DBLP:conf/icga/JonesF95,DBLP:series/sci/PitzerA12} and those are related to SAS planning (e.g., those in \textbf{RQ4}). Statistical significance is also measured. However, we acknowledge that examining more metrics for the properties of landscapes and using alternative baselines may reveal more insights, which we will plan to do in future work.


The SASs and environments studied may be subject to the threats of \textbf{external validity}. We mitigated this by using four commonly studied SAS that is of different domains, scales, and performance attributes, together with 14 environments, as used in prior work~\cite{ChenMMO21,nair2018finding,DBLP:conf/mascots/JamshidiC16,DBLP:conf/mascots/MendesCRG20,DBLP:conf/sigsoft/JamshidiVKS18}. Nonetheless, we agree that studying additional systems/environments, even other types of SASs, may prove fruitful.

\section{Related Work}
\label{sec:related}

We now discuss the work related to the landscape analysis for SASs.

\textbf{\textit{Planning for SASs:}} Over the last decade, various planning algorithms have been proposed/adopted for SASs, including the ones that rely on exact~\cite{DBLP:conf/sigsoft/ElkhodaryEM10,DBLP:journals/tse/CalinescuGKMT11,DBLP:conf/icse/Kumar0BB20,DBLP:conf/icse/ChenB14} and stochastic planning~\cite{DBLP:conf/icac/RamirezKCM09,DBLP:conf/sigsoft/ShahbazianKBM20,Chen2018FEMOSAA,DBLP:journals/taas/KinneerGG21}. However, those planners were designed under certain hypotheses instead of understanding/evidence of the planning landscapes. We advocate that future planner design for SASs should be evidence-driven, supported by not only hypotheses but also clear insights about the properties and characteristics of the planning landscape~\cite{DBLP:journals/pieee/ChenBY20} --- the key point that this paper trying to make.

\textbf{\textit{Landscape Analysis for SASs and Configurable Systems:}} Landscape analysis is a fairly new topic for configurable systems and SASs. Jamshidi and Casale~\cite{DBLP:conf/mascots/JamshidiC16} have briefly showcased the local optima of a configurable system, but this is achieved via visualizing the projected landscape, which does not provide a comprehensive summary. Likewise, Donckt \textit{et al.}~\cite{DBLP:conf/icse/DoncktWQDM20} also contribute to a simple analysis via 3D visualization. Fredericks \textit{et al.}~\cite{DBLP:conf/saso/FredericksGK019} empirically compare the planners, but they do not comment about the landscape. Recently, Li \textit{et al.}~\cite{DBLP:journals/corr/abs-2201-01429} apply local optima network --- a special type of visualization graph --- to study the landscape of configurable system. Our work differs from the above in that we demonstrate how quantifiable metrics and notions from the domain of fitness landscape analysis can be used to study SASs with respect to different environments while revealing interesting insights.



\textbf{\textit{Performance Analysis for SASs and Configurable Systems:}}  Another different but related topic is performance analysis for SASs and configurable systems, which concerns modeling the correlation between adaptation plan and performance, such as Jamshidi \textit{et al.}~\cite{DBLP:conf/kbse/JamshidiSVKPA17} and Chen~\cite{DBLP:conf/icse/Chen19b}. Those studies are orthogonal to this work as they are complementary to each other. For example, Jamshidi \textit{et al.}~\cite{DBLP:conf/kbse/JamshidiSVKPA17} state that the model can be linearly transferred between workloads (the environment in this work). This matches with our finding that there is often a significant overlap on global/local optima of the planning landscapes between environments (\textbf{RQ4}) --- a possible explanation on why they are linearly transferrable. 



\section{Conclusion and Future Work}

\label{sec:con}

In this paper, we demonstrate how the metrics and notions from the domain of fitness landscape analysis can be derived for analyzing the planning landscapes of SASs. We study four real-wold SASs under 14 different environments. Our results reveal several findings that can hint on the future research for SAS planner design.

We hope that this work can serve as a good starting point to raise the importance of planning landscape analysis for SASs and spark a dialog on a set of relevant future research directions for SAS planning. As such, the next stage on this research thread is vast, including building a dedicated methodology for SASs and fully exploring the implications as we discussed in the paper.

\balance
\bibliographystyle{ACM-Reference-Format}
\bibliography{reference} 


\begin{thebibliography}{49}


\ifx \showCODEN    \undefined \def \showCODEN     #1{\unskip}     \fi
\ifx \showDOI      \undefined \def \showDOI       #1{#1}\fi
\ifx \showISBNx    \undefined \def \showISBNx     #1{\unskip}     \fi
\ifx \showISBNxiii \undefined \def \showISBNxiii  #1{\unskip}     \fi
\ifx \showISSN     \undefined \def \showISSN      #1{\unskip}     \fi
\ifx \showLCCN     \undefined \def \showLCCN      #1{\unskip}     \fi
\ifx \shownote     \undefined \def \shownote      #1{#1}          \fi
\ifx \showarticletitle \undefined \def \showarticletitle #1{#1}   \fi
\ifx \showURL      \undefined \def \showURL       {\relax}        \fi
\providecommand\bibfield[2]{#2}
\providecommand\bibinfo[2]{#2}
\providecommand\natexlab[1]{#1}
\providecommand\showeprint[2][]{arXiv:#2}

\bibitem[\protect\citeauthoryear{Anderson, Walmsley, and Patros}{Anderson
  et~al\mbox{.}}{2021}]%
        {DBLP:conf/acsos/AndersonWP21}
\bibfield{author}{\bibinfo{person}{Chris Anderson},
  \bibinfo{person}{Timothy~Gordon Walmsley}, {and} \bibinfo{person}{Panos
  Patros}.} \bibinfo{year}{2021}\natexlab{}.
\newblock \showarticletitle{A Self-Learning Architecture for Digital Twins with
  Self-Protection}. In \bibinfo{booktitle}{\emph{{IEEE} International
  Conference on Autonomic Computing and Self-Organizing Systems, {ACSOS} 2021,
  Washington, DC, USA, September 27 - Oct. 1, 2021}}.
  \bibinfo{publisher}{{IEEE}}, \bibinfo{pages}{291--292}.
\newblock


\bibitem[\protect\citeauthoryear{Calinescu, Grunske, Kwiatkowska, Mirandola,
  and Tamburrelli}{Calinescu et~al\mbox{.}}{2011}]%
        {DBLP:journals/tse/CalinescuGKMT11}
\bibfield{author}{\bibinfo{person}{Radu Calinescu}, \bibinfo{person}{Lars
  Grunske}, \bibinfo{person}{Marta~Z. Kwiatkowska}, \bibinfo{person}{Raffaela
  Mirandola}, {and} \bibinfo{person}{Giordano Tamburrelli}.}
  \bibinfo{year}{2011}\natexlab{}.
\newblock \showarticletitle{Dynamic QoS Management and Optimization in
  Service-Based Systems}.
\newblock \bibinfo{journal}{\emph{{IEEE} Trans. Software Eng.}}
  \bibinfo{volume}{37}, \bibinfo{number}{3} (\bibinfo{year}{2011}),
  \bibinfo{pages}{387--409}.
\newblock


\bibitem[\protect\citeauthoryear{C{\'{a}}mara, Papadopoulos, Vogel, Weyns,
  Garlan, Huang, and Tei}{C{\'{a}}mara et~al\mbox{.}}{2020}]%
        {DBLP:conf/icse/CamaraP0WGHT20}
\bibfield{author}{\bibinfo{person}{Javier C{\'{a}}mara},
  \bibinfo{person}{Alessandro~Vittorio Papadopoulos}, \bibinfo{person}{Thomas
  Vogel}, \bibinfo{person}{Danny Weyns}, \bibinfo{person}{David Garlan},
  \bibinfo{person}{Shihong Huang}, {and} \bibinfo{person}{Kenji Tei}.}
  \bibinfo{year}{2020}\natexlab{}.
\newblock \showarticletitle{Towards bridging the gap between control and
  self-adaptive system properties}. In \bibinfo{booktitle}{\emph{{SEAMS} '20:
  {IEEE/ACM} 15th International Symposium on Software Engineering for Adaptive
  and Self-Managing Systems, Seoul, Republic of Korea, 29 June - 3 July,
  2020}}. \bibinfo{publisher}{{ACM}}, \bibinfo{pages}{78--84}.
\newblock


\bibitem[\protect\citeauthoryear{Chen}{Chen}{2019}]%
        {DBLP:conf/icse/Chen19b}
\bibfield{author}{\bibinfo{person}{Tao Chen}.} \bibinfo{year}{2019}\natexlab{}.
\newblock \showarticletitle{All versus one: an empirical comparison on
  retrained and incremental machine learning for modeling performance of
  adaptable software}. In \bibinfo{booktitle}{\emph{Proceedings of the 14th
  International Symposium on Software Engineering for Adaptive and
  Self-Managing Systems, May 25-31, 2019}}. \bibinfo{publisher}{{ACM}},
  \bibinfo{pages}{157--168}.
\newblock


\bibitem[\protect\citeauthoryear{Chen}{Chen}{2022}]%
        {ChenLiDOS22}
\bibfield{author}{\bibinfo{person}{Tao Chen}.} \bibinfo{year}{2022}\natexlab{}.
\newblock \showarticletitle{Lifelong dynamic optimization for self-adaptive
  systems: fact or fiction?}. In \bibinfo{booktitle}{\emph{{SANER} '22: 29th
  {IEEE} International Conference on Software Analysis, Evolution and
  Reengineering, Hawaii, United States, March 15-18 2022}}.
  \bibinfo{publisher}{{IEEE}}.
\newblock


\bibitem[\protect\citeauthoryear{Chen and Bahsoon}{Chen and Bahsoon}{2013}]%
        {DBLP:conf/icse/ChenB13}
\bibfield{author}{\bibinfo{person}{Tao Chen} {and} \bibinfo{person}{Rami
  Bahsoon}.} \bibinfo{year}{2013}\natexlab{}.
\newblock \showarticletitle{Self-adaptive and sensitivity-aware QoS modeling
  for the cloud}. In \bibinfo{booktitle}{\emph{Proceedings of the 8th
  International Symposium on Software Engineering for Adaptive and
  Self-Managing Systems, {SEAMS} 2013, San Francisco, CA, USA, May 20-21,
  2013}}. \bibinfo{publisher}{{IEEE} Computer Society},
  \bibinfo{pages}{43--52}.
\newblock


\bibitem[\protect\citeauthoryear{Chen and Bahsoon}{Chen and Bahsoon}{2014}]%
        {DBLP:conf/icse/ChenB14}
\bibfield{author}{\bibinfo{person}{Tao Chen} {and} \bibinfo{person}{Rami
  Bahsoon}.} \bibinfo{year}{2014}\natexlab{}.
\newblock \showarticletitle{Symbiotic and sensitivity-aware architecture for
  globally-optimal benefit in self-adaptive cloud}. In
  \bibinfo{booktitle}{\emph{9th International Symposium on Software Engineering
  for Adaptive and Self-Managing Systems, {SEAMS} 2014, Proceedings, Hyderabad,
  India, June 2-3, 2014}}. \bibinfo{publisher}{{ACM}}, \bibinfo{pages}{85--94}.
\newblock


\bibitem[\protect\citeauthoryear{Chen and Bahsoon}{Chen and Bahsoon}{2017a}]%
        {DBLP:journals/tse/ChenB17}
\bibfield{author}{\bibinfo{person}{Tao Chen} {and} \bibinfo{person}{Rami
  Bahsoon}.} \bibinfo{year}{2017}\natexlab{a}.
\newblock \showarticletitle{Self-Adaptive and Online QoS Modeling for
  Cloud-Based Software Services}.
\newblock \bibinfo{journal}{\emph{{IEEE} Transactions on Software
  Engineering.}} \bibinfo{volume}{43}, \bibinfo{number}{5}
  (\bibinfo{year}{2017}), \bibinfo{pages}{453--475}.
\newblock
\urldef\tempurl%
\url{https://doi.org/10.1109/TSE.2016.2608826}
\showDOI{\tempurl}


\bibitem[\protect\citeauthoryear{Chen and Bahsoon}{Chen and Bahsoon}{2017b}]%
        {DBLP:journals/tsc/ChenB17}
\bibfield{author}{\bibinfo{person}{Tao Chen} {and} \bibinfo{person}{Rami
  Bahsoon}.} \bibinfo{year}{2017}\natexlab{b}.
\newblock \showarticletitle{Self-Adaptive Trade-off Decision Making for
  Autoscaling Cloud-Based Services}.
\newblock \bibinfo{journal}{\emph{{IEEE} Transactions on Services Computing.}}
  \bibinfo{volume}{10}, \bibinfo{number}{4} (\bibinfo{year}{2017}),
  \bibinfo{pages}{618--632}.
\newblock
\urldef\tempurl%
\url{https://doi.org/10.1109/TSC.2015.2499770}
\showDOI{\tempurl}


\bibitem[\protect\citeauthoryear{Chen, Bahsoon, and Yao}{Chen
  et~al\mbox{.}}{2018a}]%
        {DBLP:journals/csur/ChenBY18}
\bibfield{author}{\bibinfo{person}{Tao Chen}, \bibinfo{person}{Rami Bahsoon},
  {and} \bibinfo{person}{Xin Yao}.} \bibinfo{year}{2018}\natexlab{a}.
\newblock \showarticletitle{A Survey and Taxonomy of Self-Aware and
  Self-Adaptive Cloud Autoscaling Systems}.
\newblock \bibinfo{journal}{\emph{{ACM} Computing Survey}}
  \bibinfo{volume}{51}, \bibinfo{number}{3} (\bibinfo{year}{2018}),
  \bibinfo{pages}{61:1--61:40}.
\newblock
\urldef\tempurl%
\url{https://doi.org/10.1145/3190507}
\showDOI{\tempurl}


\bibitem[\protect\citeauthoryear{Chen, Bahsoon, and Yao}{Chen
  et~al\mbox{.}}{2020}]%
        {DBLP:journals/pieee/ChenBY20}
\bibfield{author}{\bibinfo{person}{Tao Chen}, \bibinfo{person}{Rami Bahsoon},
  {and} \bibinfo{person}{Xin Yao}.} \bibinfo{year}{2020}\natexlab{}.
\newblock \showarticletitle{Synergizing Domain Expertise With Self-Awareness in
  Software Systems: {A} Patternized Architecture Guideline}.
\newblock \bibinfo{journal}{\emph{Proc. {IEEE}}} \bibinfo{volume}{108},
  \bibinfo{number}{7} (\bibinfo{year}{2020}), \bibinfo{pages}{1094--1126}.
\newblock


\bibitem[\protect\citeauthoryear{Chen, Li, Bahsoon, and Yao}{Chen
  et~al\mbox{.}}{2018c}]%
        {Chen2018FEMOSAA}
\bibfield{author}{\bibinfo{person}{Tao Chen}, \bibinfo{person}{Ke Li},
  \bibinfo{person}{Rami Bahsoon}, {and} \bibinfo{person}{Xin Yao}.}
  \bibinfo{year}{2018}\natexlab{c}.
\newblock \showarticletitle{{FEMOSAA}: Feature Guided and Knee Driven
  Multi-Objective Optimization for Self-Adaptive Software}.
\newblock \bibinfo{journal}{\emph{ACM Transactions on Software Engineering and
  Methodology}} \bibinfo{volume}{27}, \bibinfo{number}{2}
  (\bibinfo{year}{2018}).
\newblock


\bibitem[\protect\citeauthoryear{Chen and Li}{Chen and Li}{2021a}]%
        {DBLP:journals/corr/abs-2112-07303}
\bibfield{author}{\bibinfo{person}{Tao Chen} {and} \bibinfo{person}{Miqing
  Li}.} \bibinfo{year}{2021}\natexlab{a}.
\newblock \showarticletitle{{MMO:} Meta Multi-Objectivization for Software
  Configuration Tuning}.
\newblock \bibinfo{journal}{\emph{CoRR}}  \bibinfo{volume}{abs/2112.07303}
  (\bibinfo{year}{2021}).
\newblock
\showeprint[arXiv]{2112.07303}


\bibitem[\protect\citeauthoryear{Chen and Li}{Chen and Li}{2021b}]%
        {ChenMMO21}
\bibfield{author}{\bibinfo{person}{Tao Chen} {and} \bibinfo{person}{Miqing
  Li}.} \bibinfo{year}{2021}\natexlab{b}.
\newblock \showarticletitle{Multi-objectivizing software configuration tuning}.
  In \bibinfo{booktitle}{\emph{{ESEC/FSE} '21: 29th {ACM} Joint European
  Software Engineering Conference and Symposium on the Foundations of Software
  Engineering, Athens, Greece, August 23-28, 2021}}.
  \bibinfo{publisher}{{ACM}}, \bibinfo{pages}{453--465}.
\newblock
\urldef\tempurl%
\url{https://doi.org/10.1145/3468264.3468555}
\showDOI{\tempurl}


\bibitem[\protect\citeauthoryear{Chen, Li, and Yao}{Chen
  et~al\mbox{.}}{2018b}]%
        {DBLP:conf/gecco/0001LY18}
\bibfield{author}{\bibinfo{person}{Tao Chen}, \bibinfo{person}{Miqing Li},
  {and} \bibinfo{person}{Xin Yao}.} \bibinfo{year}{2018}\natexlab{b}.
\newblock \showarticletitle{On the effects of seeding strategies: a case for
  search-based multi-objective service composition}. In
  \bibinfo{booktitle}{\emph{the Genetic and Evolutionary Computation
  Conference, July 15-19, 2018}}. \bibinfo{publisher}{{ACM}},
  \bibinfo{pages}{1419--1426}.
\newblock


\bibitem[\protect\citeauthoryear{Chen, Li, and Yao}{Chen et~al\mbox{.}}{2019}]%
        {DBLP:journals/infsof/ChenLY19}
\bibfield{author}{\bibinfo{person}{Tao Chen}, \bibinfo{person}{Miqing Li},
  {and} \bibinfo{person}{Xin Yao}.} \bibinfo{year}{2019}\natexlab{}.
\newblock \showarticletitle{Standing on the shoulders of giants: Seeding
  search-based multi-objective optimization with prior knowledge for software
  service composition}.
\newblock \bibinfo{journal}{\emph{Info. Software Technology}}
  \bibinfo{volume}{114} (\bibinfo{year}{2019}), \bibinfo{pages}{155--175}.
\newblock


\bibitem[\protect\citeauthoryear{Ding, Liu, and Qian}{Ding
  et~al\mbox{.}}{2015}]%
        {DBLP:conf/icpads/DingLQ15}
\bibfield{author}{\bibinfo{person}{Xiaoan Ding}, \bibinfo{person}{Yi Liu},
  {and} \bibinfo{person}{Depei Qian}.} \bibinfo{year}{2015}\natexlab{}.
\newblock \showarticletitle{JellyFish: Online Performance Tuning with Adaptive
  Configuration and Elastic Container in Hadoop Yarn}. In
  \bibinfo{booktitle}{\emph{21st {IEEE} International Conference on Parallel
  and Distributed Systems, {ICPADS} 2015, Melbourne, Australia, December 14-17,
  2015}}. \bibinfo{publisher}{{IEEE} Computer Society},
  \bibinfo{pages}{831--836}.
\newblock


\bibitem[\protect\citeauthoryear{Donckt, Weyns, Quin, Donckt, and
  Michiels}{Donckt et~al\mbox{.}}{2020}]%
        {DBLP:conf/icse/DoncktWQDM20}
\bibfield{author}{\bibinfo{person}{M.~Jeroen Van~Der Donckt},
  \bibinfo{person}{Danny Weyns}, \bibinfo{person}{Federico Quin},
  \bibinfo{person}{Jonas Van~Der Donckt}, {and} \bibinfo{person}{Sam
  Michiels}.} \bibinfo{year}{2020}\natexlab{}.
\newblock \showarticletitle{Applying deep learning to reduce large adaptation
  spaces of self-adaptive systems with multiple types of goals}. In
  \bibinfo{booktitle}{\emph{{IEEE/ACM} 15th International Symposium on Software
  Engineering for Adaptive and Self-Managing Systems, Seoul, Republic of Korea,
  29 June - 3 July, 2020}}. \bibinfo{publisher}{{ACM}},
  \bibinfo{pages}{20--30}.
\newblock


\bibitem[\protect\citeauthoryear{Elkhodary, Esfahani, and Malek}{Elkhodary
  et~al\mbox{.}}{2010}]%
        {DBLP:conf/sigsoft/ElkhodaryEM10}
\bibfield{author}{\bibinfo{person}{Ahmed~M. Elkhodary}, \bibinfo{person}{Naeem
  Esfahani}, {and} \bibinfo{person}{Sam Malek}.}
  \bibinfo{year}{2010}\natexlab{}.
\newblock \showarticletitle{{FUSION:} a framework for engineering self-tuning
  self-adaptive software systems}. In \bibinfo{booktitle}{\emph{Proceedings of
  the 18th {ACM} {SIGSOFT} International Symposium on Foundations of Software
  Engineering, 2010, Santa Fe, NM, USA, November 7-11, 2010}}.
  \bibinfo{publisher}{{ACM}}, \bibinfo{pages}{7--16}.
\newblock


\bibitem[\protect\citeauthoryear{Fisher}{Fisher}{1915}]%
        {fisher1915frequency}
\bibfield{author}{\bibinfo{person}{Ronald~A Fisher}.}
  \bibinfo{year}{1915}\natexlab{}.
\newblock \showarticletitle{Frequency distribution of the values of the
  correlation coefficient in samples from an indefinitely large population}.
\newblock \bibinfo{journal}{\emph{Biometrika}} \bibinfo{volume}{10},
  \bibinfo{number}{4} (\bibinfo{year}{1915}), \bibinfo{pages}{507--521}.
\newblock


\bibitem[\protect\citeauthoryear{Fredericks, Gerostathopoulos, Krupitzer, and
  Vogel}{Fredericks et~al\mbox{.}}{2019}]%
        {DBLP:conf/saso/FredericksGK019}
\bibfield{author}{\bibinfo{person}{Erik~M. Fredericks}, \bibinfo{person}{Ilias
  Gerostathopoulos}, \bibinfo{person}{Christian Krupitzer}, {and}
  \bibinfo{person}{Thomas Vogel}.} \bibinfo{year}{2019}\natexlab{}.
\newblock \showarticletitle{Planning as Optimization: Dynamically Discovering
  Optimal Configurations for Runtime Situations}. In
  \bibinfo{booktitle}{\emph{13th {IEEE} International Conference on
  Self-Adaptive and Self-Organizing Systems, June 16-20, 2019}}.
  \bibinfo{publisher}{{IEEE}}, \bibinfo{pages}{1--10}.
\newblock


\bibitem[\protect\citeauthoryear{Gerasimou, Calinescu, and
  Tamburrelli}{Gerasimou et~al\mbox{.}}{2018}]%
        {DBLP:journals/ase/GerasimouCT18}
\bibfield{author}{\bibinfo{person}{Simos Gerasimou}, \bibinfo{person}{Radu
  Calinescu}, {and} \bibinfo{person}{Giordano Tamburrelli}.}
  \bibinfo{year}{2018}\natexlab{}.
\newblock \showarticletitle{Synthesis of probabilistic models for
  quality-of-service software engineering}.
\newblock \bibinfo{journal}{\emph{Autom. Softw. Eng.}} \bibinfo{volume}{25},
  \bibinfo{number}{4} (\bibinfo{year}{2018}), \bibinfo{pages}{785--831}.
\newblock
\urldef\tempurl%
\url{https://doi.org/10.1007/s10515-018-0235-8}
\showDOI{\tempurl}


\bibitem[\protect\citeauthoryear{Horn and Goldberg}{Horn and Goldberg}{1994}]%
        {DBLP:conf/foga/HornG94}
\bibfield{author}{\bibinfo{person}{Jeffrey Horn} {and}
  \bibinfo{person}{David~E. Goldberg}.} \bibinfo{year}{1994}\natexlab{}.
\newblock \showarticletitle{Genetic Algorithm Difficulty and the Modality of
  Fitness Landscapes}. In \bibinfo{booktitle}{\emph{Proceedings of the Third
  Workshop on Foundations of Genetic Algorithms. July 31 - August 2 1994}}.
  \bibinfo{pages}{243--269}.
\newblock


\bibitem[\protect\citeauthoryear{Jamshidi and Casale}{Jamshidi and
  Casale}{2016}]%
        {DBLP:conf/mascots/JamshidiC16}
\bibfield{author}{\bibinfo{person}{Pooyan Jamshidi} {and}
  \bibinfo{person}{Giuliano Casale}.} \bibinfo{year}{2016}\natexlab{}.
\newblock \showarticletitle{An Uncertainty-Aware Approach to Optimal
  Configuration of Stream Processing Systems}. In
  \bibinfo{booktitle}{\emph{24th {IEEE} International Symposium on Modeling,
  Analysis and Simulation of Computer and Telecommunication Systems, London,
  United Kingdom, September 19-21, 2016}}. \bibinfo{publisher}{{IEEE}},
  \bibinfo{pages}{39--48}.
\newblock


\bibitem[\protect\citeauthoryear{Jamshidi, Siegmund, Velez, K{\"{a}}stner,
  Patel, and Agarwal}{Jamshidi et~al\mbox{.}}{2017}]%
        {DBLP:conf/kbse/JamshidiSVKPA17}
\bibfield{author}{\bibinfo{person}{Pooyan Jamshidi}, \bibinfo{person}{Norbert
  Siegmund}, \bibinfo{person}{Miguel Velez}, \bibinfo{person}{Christian
  K{\"{a}}stner}, \bibinfo{person}{Akshay Patel}, {and} \bibinfo{person}{Yuvraj
  Agarwal}.} \bibinfo{year}{2017}\natexlab{}.
\newblock \showarticletitle{Transfer learning for performance modeling of
  configurable systems: an exploratory analysis}. In
  \bibinfo{booktitle}{\emph{Proceedings of the 32nd {IEEE/ACM} International
  Conference on Automated Software Engineering, {ASE} 2017, Urbana, IL, USA,
  October 30 - November 03, 2017}}. \bibinfo{publisher}{{IEEE} Computer
  Society}, \bibinfo{pages}{497--508}.
\newblock


\bibitem[\protect\citeauthoryear{Jamshidi, Velez, K{\"{a}}stner, and
  Siegmund}{Jamshidi et~al\mbox{.}}{2018}]%
        {DBLP:conf/sigsoft/JamshidiVKS18}
\bibfield{author}{\bibinfo{person}{Pooyan Jamshidi}, \bibinfo{person}{Miguel
  Velez}, \bibinfo{person}{Christian K{\"{a}}stner}, {and}
  \bibinfo{person}{Norbert Siegmund}.} \bibinfo{year}{2018}\natexlab{}.
\newblock \showarticletitle{Learning to sample: exploiting similarities across
  environments to learn performance models for configurable systems}. In
  \bibinfo{booktitle}{\emph{Proceedings of the 2018 {ACM} Joint Meeting on
  European Software Engineering Conference and Symposium on the Foundations of
  Software Engineering, November 04-09, 2018}}. \bibinfo{publisher}{{ACM}},
  \bibinfo{pages}{71--82}.
\newblock


\bibitem[\protect\citeauthoryear{Jones and Forrest}{Jones and Forrest}{1995}]%
        {DBLP:conf/icga/JonesF95}
\bibfield{author}{\bibinfo{person}{Terry Jones} {and}
  \bibinfo{person}{Stephanie Forrest}.} \bibinfo{year}{1995}\natexlab{}.
\newblock \showarticletitle{Fitness Distance Correlation as a Measure of
  Problem Difficulty for Genetic Algorithms}. In
  \bibinfo{booktitle}{\emph{Proceedings of the 6th International Conference on
  Genetic Algorithms, July 15-19, 1995}}. \bibinfo{pages}{184--192}.
\newblock


\bibitem[\protect\citeauthoryear{Kinneer, Garlan, and Goues}{Kinneer
  et~al\mbox{.}}{2021}]%
        {DBLP:journals/taas/KinneerGG21}
\bibfield{author}{\bibinfo{person}{Cody Kinneer}, \bibinfo{person}{David
  Garlan}, {and} \bibinfo{person}{Claire~Le Goues}.}
  \bibinfo{year}{2021}\natexlab{}.
\newblock \showarticletitle{Information Reuse and Stochastic Search: Managing
  Uncertainty in Self-* Systems}.
\newblock \bibinfo{journal}{\emph{{ACM} Trans. Auton. Adapt. Syst.}}
  \bibinfo{volume}{15}, \bibinfo{number}{1} (\bibinfo{year}{2021}),
  \bibinfo{pages}{3:1--3:36}.
\newblock
\urldef\tempurl%
\url{https://doi.org/10.1145/3440119}
\showDOI{\tempurl}


\bibitem[\protect\citeauthoryear{Kumar, Chen, Bahsoon, and Buyya}{Kumar
  et~al\mbox{.}}{2020}]%
        {DBLP:conf/icse/Kumar0BB20}
\bibfield{author}{\bibinfo{person}{Satish Kumar}, \bibinfo{person}{Tao Chen},
  \bibinfo{person}{Rami Bahsoon}, {and} \bibinfo{person}{Rajkumar Buyya}.}
  \bibinfo{year}{2020}\natexlab{}.
\newblock \showarticletitle{{DATESSO:} self-adapting service composition with
  debt-aware two levels constraint reasoning}. In
  \bibinfo{booktitle}{\emph{{IEEE/ACM} 15th International Symposium on Software
  Engineering for Adaptive and Self-Managing Systems, 29 June - 3 July, 2020}}.
  \bibinfo{publisher}{{ACM}}, \bibinfo{pages}{96--107}.
\newblock


\bibitem[\protect\citeauthoryear{Lesoil, Acher, T{\"{e}}rnava, Blouin, and
  J{\'{e}}z{\'{e}}quel}{Lesoil et~al\mbox{.}}{2021}]%
        {DBLP:conf/splc/LesoilATBJ21}
\bibfield{author}{\bibinfo{person}{Luc Lesoil}, \bibinfo{person}{Mathieu
  Acher}, \bibinfo{person}{Xhevahire T{\"{e}}rnava}, \bibinfo{person}{Arnaud
  Blouin}, {and} \bibinfo{person}{Jean{-}Marc J{\'{e}}z{\'{e}}quel}.}
  \bibinfo{year}{2021}\natexlab{}.
\newblock \showarticletitle{The interplay of compile-time and run-time options
  for performance prediction}. In \bibinfo{booktitle}{\emph{{SPLC} '21: 25th
  {ACM} International Systems and Software Product Line Conference, September
  6-11, 2021}}. \bibinfo{publisher}{{ACM}}, \bibinfo{pages}{100--111}.
\newblock


\bibitem[\protect\citeauthoryear{Li, Mao, and Chen}{Li et~al\mbox{.}}{2022}]%
        {DBLP:journals/corr/abs-2201-01429}
\bibfield{author}{\bibinfo{person}{Ke Li}, \bibinfo{person}{Peili Mao}, {and}
  \bibinfo{person}{Tao Chen}.} \bibinfo{year}{2022}\natexlab{}.
\newblock \showarticletitle{LONViZ: Unboxing the black-box of Configurable
  Software Systems from a Complex Networks Perspective}.
\newblock \bibinfo{journal}{\emph{CoRR}}  \bibinfo{volume}{abs/2201.01429}
  (\bibinfo{year}{2022}).
\newblock
\showeprint[arXiv]{2201.01429}
\urldef\tempurl%
\url{https://arxiv.org/abs/2201.01429}
\showURL{%
\tempurl}


\bibitem[\protect\citeauthoryear{Li, Xiang, Chen, and Tan}{Li
  et~al\mbox{.}}{2020a}]%
        {DBLP:conf/kbse/LiXCT20}
\bibfield{author}{\bibinfo{person}{Ke Li}, \bibinfo{person}{Zilin Xiang},
  \bibinfo{person}{Tao Chen}, {and} \bibinfo{person}{Kay~Chen Tan}.}
  \bibinfo{year}{2020}\natexlab{a}.
\newblock \showarticletitle{BiLO-CPDP: Bi-Level Programming for Automated Model
  Discovery in Cross-Project Defect Prediction}. In
  \bibinfo{booktitle}{\emph{35th {IEEE/ACM} International Conference on
  Automated Software Engineering, {ASE} 2020, Melbourne, Australia, September
  21-25, 2020}}. \bibinfo{publisher}{{IEEE}}, \bibinfo{pages}{573--584}.
\newblock


\bibitem[\protect\citeauthoryear{Li, Xiang, Chen, Wang, and Tan}{Li
  et~al\mbox{.}}{2020b}]%
        {DBLP:conf/icse/LiX0WT20}
\bibfield{author}{\bibinfo{person}{Ke Li}, \bibinfo{person}{Zilin Xiang},
  \bibinfo{person}{Tao Chen}, \bibinfo{person}{Shuo Wang}, {and}
  \bibinfo{person}{Kay~Chen Tan}.} \bibinfo{year}{2020}\natexlab{b}.
\newblock \showarticletitle{Understanding the automated parameter optimization
  on transfer learning for cross-project defect prediction: an empirical
  study}. In \bibinfo{booktitle}{\emph{{ICSE} '20: 42nd International
  Conference on Software Engineering, Seoul, South Korea, 27 June - 19 July,
  2020}}. \bibinfo{publisher}{{ACM}}, \bibinfo{pages}{566--577}.
\newblock


\bibitem[\protect\citeauthoryear{Li, Zeng, Meng, Tan, Zhang, Butt, and
  Fuller}{Li et~al\mbox{.}}{2014}]%
        {DBLP:conf/hpdc/LiZMTZBF14}
\bibfield{author}{\bibinfo{person}{Min Li}, \bibinfo{person}{Liangzhao Zeng},
  \bibinfo{person}{Shicong Meng}, \bibinfo{person}{Jian Tan},
  \bibinfo{person}{Li Zhang}, \bibinfo{person}{Ali~Raza Butt}, {and}
  \bibinfo{person}{Nicholas~C. Fuller}.} \bibinfo{year}{2014}\natexlab{}.
\newblock \showarticletitle{{MRONLINE:} MapReduce online performance tuning}.
  In \bibinfo{booktitle}{\emph{The 23rd International Symposium on
  High-Performance Parallel and Distributed Computing, HPDC'14, Vancouver, BC,
  Canada - June 23 - 27, 2014}}. \bibinfo{publisher}{{ACM}},
  \bibinfo{pages}{165--176}.
\newblock


\bibitem[\protect\citeauthoryear{Mandrioli and Maggio}{Mandrioli and
  Maggio}{2020}]%
        {DBLP:conf/sigsoft/MandrioliM20}
\bibfield{author}{\bibinfo{person}{Claudio Mandrioli} {and}
  \bibinfo{person}{Martina Maggio}.} \bibinfo{year}{2020}\natexlab{}.
\newblock \showarticletitle{Testing self-adaptive software with probabilistic
  guarantees on performance metrics}. In \bibinfo{booktitle}{\emph{{ESEC/FSE}
  '20: 28th {ACM} Joint European Software Engineering Conference and Symposium
  on the Foundations of Software Engineering, Virtual Event, USA, November
  8-13, 2020}}. \bibinfo{publisher}{{ACM}}, \bibinfo{pages}{1002--1014}.
\newblock


\bibitem[\protect\citeauthoryear{Mendes, Casimiro, Romano, and Garlan}{Mendes
  et~al\mbox{.}}{2020}]%
        {DBLP:conf/mascots/MendesCRG20}
\bibfield{author}{\bibinfo{person}{Pedro Mendes}, \bibinfo{person}{Maria
  Casimiro}, \bibinfo{person}{Paolo Romano}, {and} \bibinfo{person}{David
  Garlan}.} \bibinfo{year}{2020}\natexlab{}.
\newblock \showarticletitle{TrimTuner: Efficient Optimization of Machine
  Learning Jobs in the Cloud via Sub-Sampling}. In
  \bibinfo{booktitle}{\emph{28th International Symposium on Modeling, Analysis,
  and Simulation of Computer and Telecommunication Systems, November 17-19,
  2020}}. \bibinfo{publisher}{{IEEE}}, \bibinfo{pages}{1--8}.
\newblock


\bibitem[\protect\citeauthoryear{Mersmann, Bischl, Trautmann, Preuss, Weihs,
  and Rudolph}{Mersmann et~al\mbox{.}}{2011}]%
        {DBLP:conf/gecco/MersmannBTPWR11}
\bibfield{author}{\bibinfo{person}{Olaf Mersmann}, \bibinfo{person}{Bernd
  Bischl}, \bibinfo{person}{Heike Trautmann}, \bibinfo{person}{Mike Preuss},
  \bibinfo{person}{Claus Weihs}, {and} \bibinfo{person}{G{\"{u}}nter Rudolph}.}
  \bibinfo{year}{2011}\natexlab{}.
\newblock \showarticletitle{Exploratory landscape analysis}. In
  \bibinfo{booktitle}{\emph{the Genetic and Evolutionary Computation
  Conference, July 12-16, 2011}}. \bibinfo{publisher}{{ACM}},
  \bibinfo{pages}{829--836}.
\newblock


\bibitem[\protect\citeauthoryear{Merz and Freisleben}{Merz and
  Freisleben}{2000}]%
        {DBLP:journals/tec/MerzF00}
\bibfield{author}{\bibinfo{person}{Peter Merz} {and} \bibinfo{person}{Bernd
  Freisleben}.} \bibinfo{year}{2000}\natexlab{}.
\newblock \showarticletitle{Fitness landscape analysis and memetic algorithms
  for the quadratic assignment problem}.
\newblock \bibinfo{journal}{\emph{{IEEE} Trans. Evol. Comput.}}
  \bibinfo{volume}{4}, \bibinfo{number}{4} (\bibinfo{year}{2000}),
  \bibinfo{pages}{337--352}.
\newblock
\urldef\tempurl%
\url{https://doi.org/10.1109/4235.887234}
\showDOI{\tempurl}


\bibitem[\protect\citeauthoryear{Nair, Yu, Menzies, Siegmund, and Apel}{Nair
  et~al\mbox{.}}{2020}]%
        {nair2018finding}
\bibfield{author}{\bibinfo{person}{Vivek Nair}, \bibinfo{person}{Zhe Yu},
  \bibinfo{person}{Tim Menzies}, \bibinfo{person}{Norbert Siegmund}, {and}
  \bibinfo{person}{Sven Apel}.} \bibinfo{year}{2020}\natexlab{}.
\newblock \showarticletitle{Finding faster configurations using FLASH}.
\newblock \bibinfo{journal}{\emph{{IEEE} Trans. Software Eng.}}
  \bibinfo{volume}{46}, \bibinfo{number}{7} (\bibinfo{year}{2020}).
\newblock


\bibitem[\protect\citeauthoryear{Ochoa, Qu, and Burke}{Ochoa
  et~al\mbox{.}}{2009}]%
        {DBLP:conf/gecco/OchoaQB09}
\bibfield{author}{\bibinfo{person}{Gabriela Ochoa}, \bibinfo{person}{Rong Qu},
  {and} \bibinfo{person}{Edmund~K. Burke}.} \bibinfo{year}{2009}\natexlab{}.
\newblock \showarticletitle{Analyzing the landscape of a graph based
  hyper-heuristic for timetabling problems}. In \bibinfo{booktitle}{\emph{the
  Genetic and Evolutionary Computation Conference, July 8-12, 2009}}.
  \bibinfo{publisher}{{ACM}}, \bibinfo{pages}{341--348}.
\newblock


\bibitem[\protect\citeauthoryear{Pitzer and Affenzeller}{Pitzer and
  Affenzeller}{2012}]%
        {DBLP:series/sci/PitzerA12}
\bibfield{author}{\bibinfo{person}{Erik Pitzer} {and} \bibinfo{person}{Michael
  Affenzeller}.} \bibinfo{year}{2012}\natexlab{}.
\newblock \showarticletitle{A Comprehensive Survey on Fitness Landscape
  Analysis}.
\newblock In \bibinfo{booktitle}{\emph{Recent Advances in Intelligent
  Engineering Systems}}. \bibinfo{series}{Studies in Computational
  Intelligence}, Vol.~\bibinfo{volume}{378}. \bibinfo{publisher}{Springer},
  \bibinfo{pages}{161--191}.
\newblock


\bibitem[\protect\citeauthoryear{Ramirez, Knoester, Cheng, and
  McKinley}{Ramirez et~al\mbox{.}}{2009}]%
        {DBLP:conf/icac/RamirezKCM09}
\bibfield{author}{\bibinfo{person}{Andres~J. Ramirez},
  \bibinfo{person}{David~B. Knoester}, \bibinfo{person}{Betty H.~C. Cheng},
  {and} \bibinfo{person}{Philip~K. McKinley}.} \bibinfo{year}{2009}\natexlab{}.
\newblock \showarticletitle{Applying genetic algorithms to decision making in
  autonomic computing systems}. In \bibinfo{booktitle}{\emph{Proceedings of the
  6th International Conference on Autonomic Computing, {ICAC} 2009, June 15-19,
  2009, Barcelona, Spain}}. \bibinfo{publisher}{{ACM}},
  \bibinfo{pages}{97--106}.
\newblock


\bibitem[\protect\citeauthoryear{Shahbazian, Karthik, Brun, and
  Medvidovic}{Shahbazian et~al\mbox{.}}{2020}]%
        {DBLP:conf/sigsoft/ShahbazianKBM20}
\bibfield{author}{\bibinfo{person}{Arman Shahbazian}, \bibinfo{person}{Suhrid
  Karthik}, \bibinfo{person}{Yuriy Brun}, {and} \bibinfo{person}{Nenad
  Medvidovic}.} \bibinfo{year}{2020}\natexlab{}.
\newblock \showarticletitle{eQual: informing early design decisions}. In
  \bibinfo{booktitle}{\emph{{ESEC/FSE} '20: 28th {ACM} Joint European Software
  Engineering Conference and Symposium on the Foundations of Software
  Engineering, Virtual Event, USA, November 8-13, 2020}}.
  \bibinfo{publisher}{{ACM}}, \bibinfo{pages}{1039--1051}.
\newblock


\bibitem[\protect\citeauthoryear{Siegmund, Kolesnikov, K{\"{a}}stner, Apel,
  Batory, Rosenm{\"{u}}ller, and Saake}{Siegmund et~al\mbox{.}}{2012}]%
        {DBLP:conf/icse/SiegmundKKABRS12}
\bibfield{author}{\bibinfo{person}{Norbert Siegmund},
  \bibinfo{person}{Sergiy~S. Kolesnikov}, \bibinfo{person}{Christian
  K{\"{a}}stner}, \bibinfo{person}{Sven Apel}, \bibinfo{person}{Don~S. Batory},
  \bibinfo{person}{Marko Rosenm{\"{u}}ller}, {and} \bibinfo{person}{Gunter
  Saake}.} \bibinfo{year}{2012}\natexlab{}.
\newblock \showarticletitle{Predicting performance via automated
  feature-interaction detection}. In \bibinfo{booktitle}{\emph{34th
  International Conference on Software Engineering, {ICSE} 2012, June 2-9,
  2012}}. \bibinfo{publisher}{{IEEE} Computer Society},
  \bibinfo{pages}{167--177}.
\newblock


\bibitem[\protect\citeauthoryear{Stadler}{Stadler}{1996}]%
        {stadler1996landscapes}
\bibfield{author}{\bibinfo{person}{Peter~F Stadler}.}
  \bibinfo{year}{1996}\natexlab{}.
\newblock \showarticletitle{Landscapes and their correlation functions}.
\newblock \bibinfo{journal}{\emph{Journal of Mathematical chemistry}}
  \bibinfo{volume}{20}, \bibinfo{number}{1} (\bibinfo{year}{1996}),
  \bibinfo{pages}{1--45}.
\newblock


\bibitem[\protect\citeauthoryear{Tavares, Pereira, and Costa}{Tavares
  et~al\mbox{.}}{2008}]%
        {DBLP:journals/tsmc/TavaresPC08}
\bibfield{author}{\bibinfo{person}{Jorge Tavares},
  \bibinfo{person}{Francisco~Baptista Pereira}, {and} \bibinfo{person}{Ernesto
  Costa}.} \bibinfo{year}{2008}\natexlab{}.
\newblock \showarticletitle{Multidimensional Knapsack Problem: {A} Fitness
  Landscape Analysis}.
\newblock \bibinfo{journal}{\emph{{IEEE} Trans. Syst. Man Cybern. Part {B}}}
  \bibinfo{volume}{38}, \bibinfo{number}{3} (\bibinfo{year}{2008}),
  \bibinfo{pages}{604--616}.
\newblock
\urldef\tempurl%
\url{https://doi.org/10.1109/TSMCB.2008.915539}
\showDOI{\tempurl}


\bibitem[\protect\citeauthoryear{WRIGHT}{WRIGHT}{1932}]%
        {wright1932roles}
\bibfield{author}{\bibinfo{person}{S WRIGHT}.} \bibinfo{year}{1932}\natexlab{}.
\newblock \showarticletitle{The roles of mutation, inbreeding, crossbreeding
  and selection in evolution}. In \bibinfo{booktitle}{\emph{The 6th
  international congress of Genetics}}, Vol.~\bibinfo{volume}{1}.
  \bibinfo{pages}{356--366}.
\newblock


\bibitem[\protect\citeauthoryear{Ye and Kalyanaraman}{Ye and
  Kalyanaraman}{2003}]%
        {DBLP:conf/sigmetrics/YeK03}
\bibfield{author}{\bibinfo{person}{Tao Ye} {and} \bibinfo{person}{Shivkumar
  Kalyanaraman}.} \bibinfo{year}{2003}\natexlab{}.
\newblock \showarticletitle{A recursive random search algorithm for large-scale
  network parameter configuration}. In \bibinfo{booktitle}{\emph{Proceedings of
  the International Conference on Measurements and Modeling of Computer
  Systems, {SIGMETRICS} 2003, June 9-14, 2003, San Diego, CA, {USA}}}.
  \bibinfo{publisher}{{ACM}}, \bibinfo{pages}{196--205}.
\newblock


\bibitem[\protect\citeauthoryear{Zou}{Zou}{2007}]%
        {zou2007toward}
\bibfield{author}{\bibinfo{person}{Guang~Yong Zou}.}
  \bibinfo{year}{2007}\natexlab{}.
\newblock \showarticletitle{Toward using confidence intervals to compare
  correlations.}
\newblock \bibinfo{journal}{\emph{Psychological methods}} \bibinfo{volume}{12},
  \bibinfo{number}{4} (\bibinfo{year}{2007}), \bibinfo{pages}{399}.
\newblock


\end{thebibliography}
\citestyle{acmauthoryear} 

\end{document}